\documentclass[doublecol]{epl2} 

\usepackage{graphicx,epsfig,chemarr}
\usepackage{amssymb, amsfonts, amsmath}

\def\be{\begin{equation}}
\def\ee{\end{equation}}
\def\ba{\begin{eqnarray}}
\def\ea{\end{eqnarray}}
\def\kbar{\bar{k}}
\def\Vmax{V_{\text{max}}}
\def\Vmin{V_{\text{min}}}
\def\fs{f^\star}

\title{Quorum Percolation in Living Neural Networks}

\author{Or Cohen\inst{1} \and Anna Keselman\inst{1} \and Elisha Moses\inst{1} \and Mar{\'{i}}a Rodr{\'{i}}guez Mart{\'{i}}nez\inst{1}
\and Jordi Soriano\inst{1,2} \and Tsvi Tlusty\inst{1}\thanks{Corresponding author. E-mail:
\email{tsvi.tlusty@weizmann.ac.il}}}

\shortauthor{O. Cohen \etal}

\institute{
  \inst{1} Department of Physics of Complex Systems. Weizmann Institute of Science.
Rehovot 76100, Israel.\\
  \inst{2} Departament d'ECM. Facultat de F{\'{i}}sica, Universitat de Barcelona. Av. Diagonal 647, 08028 Barcelona, Spain.\\
}


\abstract{Cooperative effects in neural networks appear because a
neuron fires only if a minimal number $m$ of its inputs are excited.
The multiple inputs requirement leads to a percolation model termed
{\it quorum percolation}. The connectivity undergoes a phase
transition as $m$ grows, from a network--spanning cluster at low $m$
to a set of disconnected clusters above a critical $m$. Both
numerical simulations and the model reproduce the experimental
results well. This allows a robust quantification of biologically
relevant quantities such as the average connectivity $\kbar$ and the
distribution of connections $p_k$.}

\begin{document}

\maketitle

Large networks generate complex behavior, as is apparent in diverse
systems such as computers, society and biology \cite{Barabasi-2002}.
In particular, the integration of information from several neighbors
leads to highly correlated and collaborative activity
\cite{Anderson-1972}. In general, the complexity and capacity of the
network increases steeply with the degree of inter--connectivity, as
does the difficulty to understand these complex networks. Biological
neural networks stand out as intriguing complex systems
\cite{Koch-1999}. From cultures in a dish \cite{Wagenaar-2006} all
the way to the brain \cite{Review-Brain}, neural networks display a
rich repertoire of activity and functionality, which arises from the
interplay between hundreds to millions of neurons. In the simplest
picture of a spiking neuron, stimuli (inputs) from connected neurons
are ``integrated'' in the target neuron, which fires once a
threshold voltage is reached and then propagates the electric signal
on to other neurons \cite{Osan-2002}.

Imposing the need for a large quorum of $m$ input nodes to fire
leads to a percolation problem, which we term ``quorum percolation''
(QP). This name also hints for the potential uses of QP to treat the
spread of diseases, rumors and opinions. In this work we show how
the multiple inputs requirement modifies significantly the function
of the network. The QP approach takes the number of inputs required
for firing $m$ as its control parameter and can then explain
phenomena such as the transition to a {\it giant $m$-connected
component} (GmCC), a continuum of connected neurons spanning a
significant fraction of the network.

The QP setting is similar to that of bootstrap percolation (BP) in
which nodes with less than $k$ neighbors are iteratively ``pruned"
from a graph until there remains only its highly interconnected
``$k$-core" (\cite{Adler-1991,Bollobas-1984,Pittel-1996,Tlusty-2009}
and references therein). The propagation of firing in QP is
analogous to the advancement of the pruning process in BP (see
\cite{Tlusty-2009}). In practice, however, physical realizations of
BP are usually limited in their connectivity, often on the order of
$\kbar=2-3$. In contrast, the average connectivity in the neural
network studied here is much higher, $\kbar=50-150$
\cite{Soriano08}, and results in a very different behavior of the
network. A variation on the problem was posed in \cite{Watts-2002},
where the propagation depends on the fraction of ignited neighbors
and was solved using the approach presented in \cite{Callaway-2000}.

\section{Model}
The neural network is modeled as a random directed graph
\cite{Breskin06} whose nodes are neurons connected by synapses. The
input degree distribution is $p_k$. Each neuron is assigned a
probability $f=f(V)$ to fire in response to the direct excitation by
an externally applied electrical voltage. The total probability of
a single neuron to fire, $\Phi$, is the sum of $f(V)$ and of its
response to the input from the firing of ``neighboring'' (i.e.
connected) neurons.  This response is characterized by the {\it
collectivity} $\Psi_m(\Phi)$, the probability that there is a quorum
of at least $m$ inputs to a neuron that fire and thus excite the
neuron.  $\Psi_m(\Phi)$ is the synaptic excitation probability,
which is the signature of collective effects in the network.
Therefore, $\Phi(m,f)$ is given by:
\begin{align}
\Phi(m,f) = f+(1-f)\Psi_m(\Phi)\hspace{2.7cm}
\nonumber \\
=f+(1-f)\sum_{k=m}^{\infty }p_{k} \sum_{l=m}^{k }  \dbinom{k}{l} \,\Phi^l \,
(1-\Phi)^{k-l}. \label{prob}
\end{align}
Equation  (\ref{prob}) represents the fixed point of an iterative
firing process that starts with the external ignition of a fraction
$f(V)$ of neurons and propagates by igniting at every step those
neurons that accumulated at least $m$ firing inputs. Equation
(\ref{prob}) exploits the effective tree--like topology of the
random network to ignore the presence of feedback loops and of
recurrent activity in the neural culture. The validity of the random
graph approximation to metric graphs such as the experimental neural
networks is discussed in detail in \cite{Tlusty-2009}. It proves
convenient to express the collectivity $\Psi_m(\Phi)$ in terms of
the generating function $J(z)$  of the degree distribution $ p_{k}$,
$J(z) = \sum_{k} \, p_{k} \, z^k$, which results in a truncated
Taylor expansion $\Psi_m(\Phi) = 1-\sum_{l=0}^{m-1 }(\Phi^l/ l!)
\partial^{\,l} J/\partial z^l |_{1-\Phi} $.

We previously \cite{Breskin06} treated a simplified version of
equation (\ref{prob}) for the standard bond percolation case of
$m=1$, and observed a percolation transition both in the model and
the experiment. The percolated phase is characterized by a giant
connected component that spans a non--zero fraction of the network.
Note that the GmCC is very different from the conventional $m = 1$
giant connected components.

Here we treat the case of an arbitrary $m$ and find how excitation
by multiple inputs changes the response of the neural network.
Within the framework of the model, all distributions of connections
p(k) in the network are possible. However, when coming to compare to
particular results in a quantitative manner the need to choose a
specific distribution arises. This allows numerical simulations of
the model and a comparison to data from the experiment. The correct
choice was determined previously \cite{Breskin06,Soriano08} to be a
Gaussian, based on numerical simulation comparing data from
experiment ---both the giant component size $g$ \cite{Soriano08} and
the cluster size distribution $p_s$ \cite{Breskin06}. After trying a
variety of distribution functions (including the natural candidates, i.e. power laws, exponential and Poisson), we concluded that the
gaussian distribution gave the best description.

Following the experiment, we specify the degree distribution $p_k$
and solve for $\Phi(m,f)$ as a function of $m$ and $f$, the
experimental control parameters. Based on our previous experimental observations \cite{Breskin06,Soriano08}, we assume a gaussian distribution $p_k \sim \text{exp}[-(k-\kbar)^2/(2\sigma^2)]$ with $\kbar=50$ and
$\sigma=15$.

\begin{figure}[hb]
\begin{center}
\includegraphics[width=7cm]{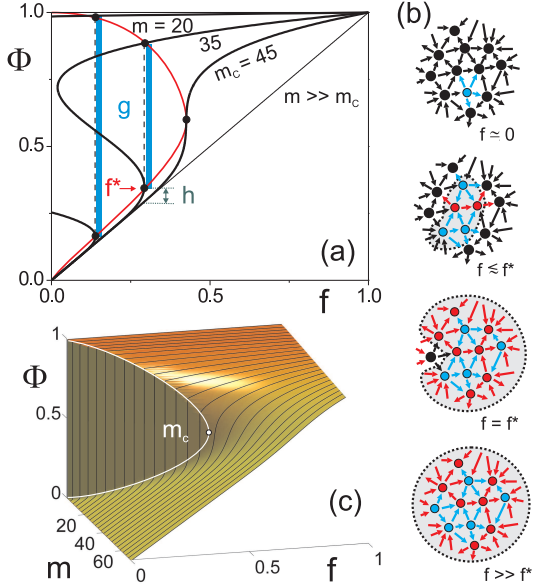} \vspace{0.0cm}
\caption{Quorum percolation model. (a) Black lines: Family of
solutions $\Phi(m,f)$ for increasing $m$ and a gaussian degree
distribution $p_k(k)$ with $\kbar=50$ and $\sigma=15$. Dashed lines
connect the physical solutions in the coexistence region. The jump
in $\Phi$ for each value of $\fs$ identifies  $g$, the size of the
GmCC (blue bars), and the contribution of the small clusters $h$.
For example, for $m=20$ at $f=0$ $\Phi$ grows continuously from zero, so
that at $f=0$ there is no GmCC. At $f\approx 0.15$ the GmCC appears,
and $\Phi$ jumps practically to $1$. While the clusters comprising
$h$ are isolated, the GmCC was observed in the simulations to be
connected. The black dots and the red line indicate the transition
points $\fs(m)$ and $m_c$. (b) Schematic representation of the
ignition of the network as a function of $f$, for a network with
$\kbar=5$ and input threshold $m=3$. Ignition occurs in two phases:
first the external ignition of a fraction $f$ of neurons and in the
second, the expansion of the firing $m$--connected cluster (GmCC).
For $f \simeq 0$ only the neurons (circles) with the lowest
excitation threshold are ignited (blue) and pass the signal (blue
arrows). As $f$ increases more neurons get excited. These neurons,
in turn, excite synaptically other neurons (red), but the activity
is confined to isolated clusters. At $f=\fs$ enough neurons fire for
the network to percolate and a GmCC emerges. Weakly connected
neurons and with high firing thresholds will fire only at large $f$.
The outlined areas show the neurons that fired together for each
$f$. (c) Three--dimensional representation of the solutions of the
model as a function of $m$ and $f$. The white curve shows the values
of the transition points $\fs$.} \label{Fig:model}
\end{center}
\end{figure}

To find the firing probability $\Phi(m,f)$ for given values of $m$
and $f$, we rearrange equation (\ref{prob}) as $f(m,\Phi) = (\Phi -
\Psi_m(\Phi))/(1 - \Psi_m(\Phi))$. In Figure \ref{Fig:model}(a),
$f(m,\Phi)$ is plotted and graphically inverted  to find the
solution $\Phi(m,f)$. The solution shows that both parameters $f$
and $m$ have transition values, $\fs(m)$ and $m_c$, where the
solution changes qualitatively. For $m<m_c$, the curve $\Phi(m,f)$
is non--monotonic and below the transition, $f<\fs(m)$, there exist
three solutions: The lower one corresponds to $\Phi\simeq f$ and
represents a system where only the externally excited neurons fire.
In this solution the external voltage $V$ is only able to ignite
'sensitive' neurons, i.e. neurons with a low excitation threshold.
After they fire, the signal dies out without propagating throughout
the network. The second, intermediate solution lies in a
non--physical region in which the firing probability decreases with
the external electrical stimulus $f$, i.e. $d\Phi/df<0$. The third,
higher solution corresponds to the synaptic excitation of most of
the neurons. This solution signals the ignition of the giant
connected component of the network. The two available physical
solutions, the un--excited network $\Phi(f)\simeq f$ and the excited
network $\Phi(f)\simeq 1$ coexist below $\fs(m)$.

The first two solutions merge and disappear at $f = \fs(m)$ [Fig.
\ref{Fig:model}(a)] and for $f>\fs(m)$ only the third solution is
available. Physically, at this point the external voltage is high
enough to excite the GmCC and the whole network percolates. As a
precursor to the transition, $\Phi(f)$ deviates from the expected
$\Phi(f)\simeq f$. This deviation signals the ignition of small
clusters $h$ that contain the most excitable neurons. At the
critical value $\fs(m)$ the system undergoes a phase transition,
from the low excitation state characterized by $\Phi\simeq f$ to the
percolation solution $\Phi\simeq 1$ where almost all neurons fire
[Fig.\ \ref{Fig:model}(b)]. The behavior of the network can be seen
as a first order phase transition that exhibits a discontinuity in
the value of $\Phi$. Note that, in contrast to the traditional $m=1$
percolation, for $m>1$ the GmCC does not appear at zero excitation.
The requirement of multiple inputs imposes a minimal excitation
$\fs(m)$, or threshold voltage, to excite the GmCC, which may
``immunize" the neural network against excitation by noise.

The size of the GmCC $g$ decreases as $m$ grows, and reaches $g=0$
at a critical value $m=m_c$. Graphically, at $f=\fs(m_{c})$ the two
physical solutions (as well as the unphysical third one) merge into
one [Fig.\ \ref{Fig:model}(a)], and for higher values of both $f$
and $m$ only one solution exists. For $m>m_c$ the demand for
multiple inputs is so restrictive that the network has no giant
connected component. The network breaks off into isolated clusters
that are ignited independently. This scenario corresponds to a
second order phase transition, characterized by the order parameter
$g$, with $g=0$ for $m > m_{c}$ and $g>0$ for $m \leq m_{c}$. At the
extreme of $m\gg m_{c}$ the network is completely disconnected,
neurons are ignited in response to the external excitation only, and
the solution $\Phi\simeq f$ applies for all values of $f$.

The non--monotonic curve of $\Phi(f)$ is similar to the van der
Waals density--pressure isotherms, which exhibit a gas--liquid
coexistence region below a critical temperature (equivalent to
$m_c$). The non--physical segment of the isotherm is resolved by
Maxwell's construction which connects the coexisting phases by a
vertical line, signifying a first order phase transition. In a
similar way, we replace the coexistence region of the $\Phi(f)$
curves by a vertical line, connecting the two physical solutions of
the system (dashed lines in Fig. \ref{Fig:model}(a)). This line
identifies the size of the GmCC $g$. In contrast to the traditional
liquid--gas transition which occurs at around the center of the
coexistence region, our simulations show that ignition of the GmCC
occurs very close to $\fs$.

Figure \ref{Fig:model}(c) summarizes the response of the network as
a function of $\Phi$, $m$ and $f$. The white curve shows $\fs(m)$,
and identifies the first order phase transition in the size of the
GmCC. This curve disappears at $m=m_c$, where the network undergoes
a second order, continuous phase transition. The point $m=m_c$ is
analogous to the classical tricritical point of thermodynamic phase
transitions.

\section{Experiment}
The ideas of QP are tested on a typical two dimensional
{\it in vitro} culture comprising of
over $5\times 10^5$ neurons. The experimental system (see
Refs. \cite{Soriano08,Breskin06,Eckmann-2007a} for details) is a
culture of neurons that are extracted from rat embryonic brains in
day $17$ or $19$ of pregnancy. Neurons are plated on a glass
coverslip and create a highly connected two--dimensional network in
about $2$ weeks. Activity can be stimulated globally in the culture
by an external voltage $V$ that is applied across two bath
electrodes.

We assume that a neuron integrates over its input connections, has a
threshold voltage $V_T$ to be ignited, and that on average each
input into a neuron contributes a voltage $g_{\text{syn}}$. Thus,
$m_0=V_T/g_{\text{syn}}$ inputs are initially required to excite a
neuron and propagate the signal in the network. The synaptic
strength can be reduced by application of a blocker (CNQX) for the
synaptic (AMPA) glutamate receptor. Administering CNQX in increasing
concentration gradually decreases the synaptic bond between neurons,
and thus increases the number of inputs $m$ needed for a neuron to
fire. The relation between the input threshold $m$ and the
concentration of CNQX is given by $m = m_0 (1+\text{[CNQX]}/K_d)$
\cite{Soriano08}, with $K_d = 300$ nM. We use $m_0 \simeq 15$ as the
the initial input threshold for the unperturbed network
\cite{Soriano08}.

Signals in the neural network are transferred within the cells by an
electric action potential, and between cells by chemicals that are
detected by specific receptors at the synapses. The major types of
receptor that a neuron has are receptors of type AMPA and NMDA for
the excitatory neurotransmitter glutamate, and receptors for the
inhibitory neurotransmitter GABA$_\text{A}$. In the experiments reported
here the NMDA and GABA$_\text{A}$ receptors were completely blocked by
their antagonists (2R)-amino-5-phosphonovaleric acid (APV) and
bicuculline respectively, leaving AMPA as the dominant receptor.
Thus inhibition is practically turned off, and we measure a network
that is uniquely excitatory.

CNQX is a highly specific antagonist of the AMPA receptors, with
little or no effect on other receptors. Its effect is to block AMPA
receptors and thus reduce the synaptic strength. This does not
directly affect the network architecture, in the sense that it does
not change any hardwired connection (although of course, for
sufficiently high concentrations the synapses are totally blocked,
and at that final stage the network topology does change).

If inhibition is not neutralized by bicuculline, then the balance of
excitatory and inhibitory neurons is crucial for the behavior of the
network. In the framework of our simplified model, it suffices
approximate the contribution of inhibitory synapses by a simple
subtraction to the membrane potential. This is based on known values
of the post synaptic current in both excitatory and inhibitory
synapses, which are similar \cite{Soriano08,Jacobi09}.

\begin{figure}[!h]
\begin{center}
\includegraphics[width=7cm]{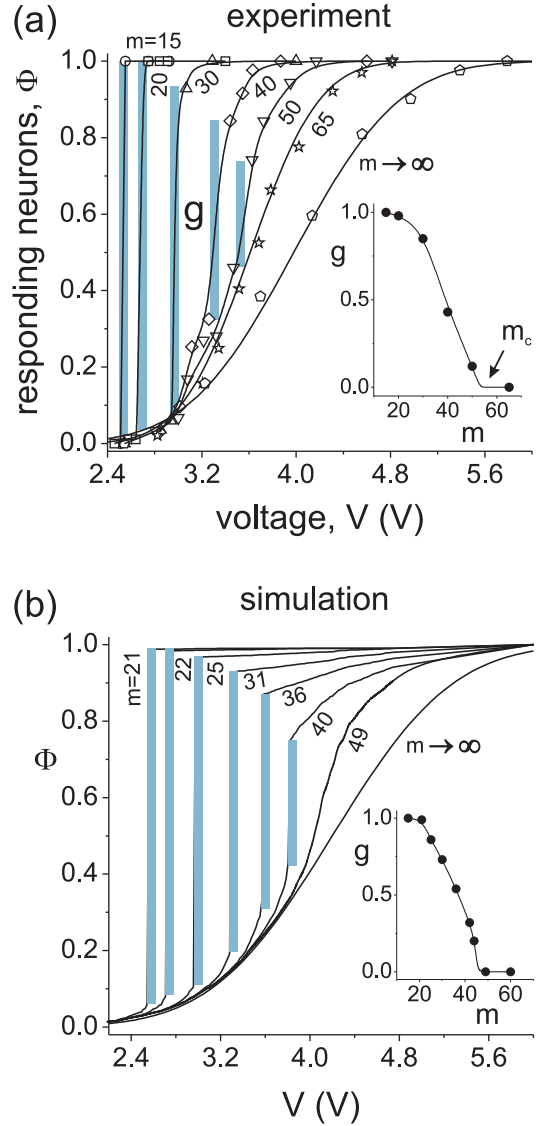} \vspace{0.0cm}
\caption{(a) Experimental $\Phi(V)$ curves for gradually higher
concentrations of CNQX, between $0$ and $10$ $\mu$M, and with $m =
m_0 (1+\text{[CNQX]}/K_d)$, where $K_d=300$ nM and $m_0=15$. (b)
$\Phi(V)$ curves from a numerical simulation of the model with a
gaussian $p_k(k)$, $\kbar=40$ and $\sigma=20$. Vertical bars show
the size of the GmCC $g$. The insets show the corresponding size of
the GmCC as a function of $m$. For the experiments, the values of $g$ are averaged over two consecutive $\Phi(V,m)$ explorations of the network. Lines are a guide to the eye.} \label{Fig:exp-simul}
\end{center}
\end{figure}

For a given network with a probability distribution of input
connections $p_k$, the pair of control parameters $(m,V)$ completely
determines the firing of the network. Initially, for the unperturbed
network, $m=m_0$ is much smaller than the average connectivity of
the neurons $\kbar=\sum_{k} kp_k$ \cite{Soriano08}. Hence, a small
fraction of firing neurons can ignite all the rest, $\Phi$ jumps abruptly to $1$ and the GmCC comprises the whole network [Fig. \ref{Fig:exp-simul}(a)]. As CNQX is added, the connectivity decreases and $m$ grows. Those neurons having less than $m$ inputs
get disconnected from the network and, in turn, reduce the number of
inputs of their target neurons. The fraction $\Phi$ of neurons that respond together to a given voltage $V$ reduces in size. The biggest jump in $\Phi$, which identifies the size $g$ of the GmCC, gradually decreases. At a critical value $m=m_c$, the GmCC disintegrates into
isolated $m$--connected clusters. When the network is fully
disconnected then the neurons respond only to the external
excitation. The value of $m_c$ is a reliable and reproducible
experimental measurement, and forms the basis for evaluating several
biologically relevant measures of the connectivity in the neural
culture \cite{Soriano08,Jacobi09}. A possibility which is ignored here but
can be accommodated in the model is that  $m$ is a function of $V$.

\section{Simulations}
A random, directed graph was created by assigning to each vertex a
number of in-- and out--edges, according to the given gaussian
distribution. The total numbers of in and out edges were equated by
randomly removing or adding out--edges. Connections were made by
iteratively connecting the pair of nodes with the highest number of
available out-- and in-- edges. The resulting graph was randomized
by switching vertices between randomly selected pairs of edges.

Each vertex was also assigned a random sensitivity $f$ to the
external voltage $V$, drawn from a gaussian distribution, which was
later fit to the experimentally measured sensitivity of the neurons
in the culture. The parameters determining each simulation were $m$
and the external voltage $V$. In a typical run $m$ was increased
from $2$ to $50$, and $V$ was varied from $\Vmin$ to $\Vmax$ so that
$f(\Vmin) \simeq 0$ and $f(\Vmax) \simeq 1$. At each $V$ those
neurons with $f(V)>0$ were first ignited, which in turn excited
connected neurons that satisfied the input threshold requirement.
The total fraction of nodes firing $\Phi(m,V)$ was stored for
further analysis. An increase of the firing response of the network
at a single step higher than $10 \%$ was considered to be the
signature of the GmCC $g(m)$. We also extracted $\fs(m)$ for each
run. All values were averaged over $10-50$ different realizations of
random graphs created with the same parameters.

\section{Comparison of Simulation and Experiment}
Figure \ref{Fig:exp-simul} shows the comparison of a particular
realization in the simulation to a single experimental run. Both
experiment and a single simulation include fluctuations in the
choice of the connections, and are therefore noisy. Still, in both
cases a finite $f$ is needed to induce the transition to a GmCC, and
the emergence of a critical $m_c$ occurs in a similar way. The inset
shows a detail of the development of the GmCC as a function of the
input threshold $m$.

\begin{figure}[!h]
\begin{center}
\includegraphics[width=7cm]{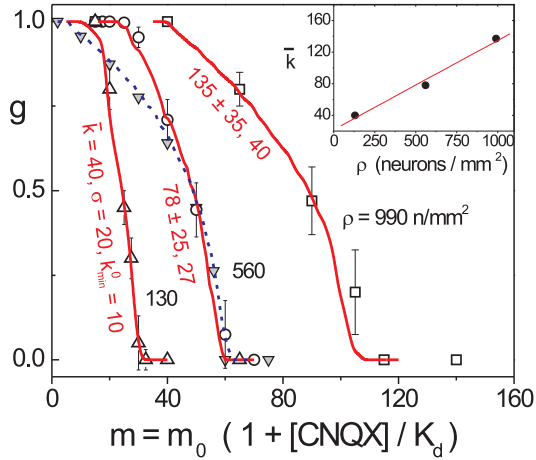} \vspace{0.0cm}
\caption{Main plot: Experimental $g(m)$ curves for $3$ different
neuronal densities (open symbols, each data set averaged over $4$
experiments) are fitted to numerical simulations of the model for
$p_k(k)$ gaussian (red, each curve averaged over $5$
realizations). The values in red indicate the parameters of the
simulations: $\kbar$, $\sigma$ and the outer cutoff
$k_{\text{min}}^o$. In all cases
$k_{\text{min}}^i=k_{\text{min}}^o$. The grey triangles show the
numerical integration of the model for $\kbar=78$ and
$\sigma=25$, compared to numerical simulations with the same
parameters and for $k_{\text{min}}^i=27$, $k_{\text{min}}^o=0$
(blue dashed line, average over $10$ realizations). Inset: average
connectivity $\kbar$ (extracted from the simulations) as a
function of the density $\rho$ of the neural culture.}
\label{Fig:fit-experiments}
\end{center}
\end{figure}

A particularly relevant comparison of experimental and simulation
results is obtained as the experimental densities of neurons is
varied. The effect of increasing the density is to increase the
average connectivity $\kbar$ that characterizes the distribution
$p_k$ \cite{Soriano08}. Figure \ref{Fig:fit-experiments} shows the
measured transition curves of $g$ as function of $m$ at different
densities. The curves are characterized by the presence of a plateau
$g(m) \simeq 1$ for low values of $m$. The GmCC gradually decreases
for higher $m$, and disappears at the transition point $m_c$.

In the simulations, the distribution $p_k$ has to be varied in order
to follow the experimental behavior. The connectivity is described
by gaussians with increasing $\kbar$ and $\sigma$. To reproduce the
experimental plateau, however, a cutoff $k_{\text{min}}^{i,o}$ in
both the input $i$ and output $o$ degree distributions has to be
applied, so that $p_k^{i,o}(k)=0$ for $k^{i,o} <
k_{\text{min}}^{i,o}$. Overall, as seen by the red lines in Fig.
\ref{Fig:fit-experiments}, the agreement of simulation and
experiment is striking, and the variation of $m_c$ is exactly
reproduced. More important, as the inset shows, the change in
experimental neuronal densities is linearly related to the changes
in average connectivity of the simulation.

\section{Comparison of Simulation and Model}
The model and simulation fit very well when using the gaussian
distribution of the connections. In a way, this is both encouraging
and surprising, since the simulations carry a large number of loops
(the number of directed triangles can be estimated by $k^3/6\sim
8\times 10^4$), while the model assumes no loops. This leads to some
discrepancies in the quantitative agreement of the simulation with
the experiment, for example for $m=49$ in the model $g=0$ while in
the experiments it is still non--zero at $g=50$. Furthermore,
measuring average quantities means that we may be describe only the
majority of the population, while some small fraction of the
neurons, too small to disrupt the average properties, may obey
statistics that deviate from the gaussian distribution.

The only sign of a deviation of model and simulation appears when we
introduce the lower cutoff in the distribution of both the input and
output connectivity. This was crucial for reproducing the plateau
that characterized $g(m)$ for low $m$ in the experiment. If the
cutoff was applied either to only the input or only to the output
distribution then the numerical simulations did not give this
plateau (dashed line in Fig. \ref{Fig:fit-experiments}). As for the
model, the cutoff in the input distribution (the model disregards
the output distribution) had practically no effect on the response
of the network (grey triangles in Fig. \ref{Fig:fit-experiments}).

\section{Discussion}The existence of a `threshold',
or the demand for a `quorum' of inputs lies at the heart of
Integrate and Fire models of the neuron \cite{Koch-Book}, and is
basically what allows for computation in neuronal ensembles
\cite{Feinerman-2008}. Similar mechanisms may be at work in diverse
systems such as
the creation of public opinion, in which people hear many views
before they make their mind up \cite{Havlin}, and where $f$ would be
construed as external forcing such as the effect of the media. We
believe that the QP model supplies a quantitative framework in which
to study such processes. In this system, the demand for ``more"
neuronal inputs leads to a ``different" transition and to a more
complex threshold behavior \cite{Feinerman-2008}.

The major departure of the QP from standard percolation is in the
ignition process. The need for a significant fraction $f^*$ of the
network to be initially activated, and the dependence of $f^*$ on
$m$, is a unique characteristic of the system. We believe that the
deviation of the simulated and experimental results from the
tree--like model occurs due to the existence of loops, which would
be important exactly in the region of small excitations that lead to
ignition of the whole network. Many loops accelerate the propagation
of the firing cluster and thus the initial fraction $f^*$ may
decrease even to zero \cite{Tlusty-2009}. The possibility of
incorporating inhibitory neurons directly in the model would also be
interesting. We furthermore expect that studies of the behavior of
$f^*(m)$ and of $m_c$ for different degree distributions will yield
a rich variety of phenomena.
 \acknowledgments

We are grateful to J.-P. Eckmann for fruitful discussions and
insight. This work was supported by the Israel Science Foundation
and by the Minerva Foundation (Munich, Germany).

\end{document}